# A New DNA Sequences Vector Space on a Genetic Code Galois Field

Robersy Sánchez [1 3*], Luis A. Perfetti[2], Ricardo Grau [2 3] and Eberto Morgado [2].

[1]Research Institute of Tropical Roots, Tuber Crops and Banana (INIVIT). Biotechnology group. Santo Domingo. Villa Clara. Cuba.

[2]Faculty of Mathematics Physics and Computation, Central University of Las Villas, Villa Clara, Cuba.

[3] Center of Studies on Informatics, Central University of Las Villas, Villa Clara, Cuba

## Abstract

A new *n*-dimensional vector space of the DNA sequences on the Galois field of the 64 codons ($GF(64)$) is proposed. In this vector space gene mutations can be considered linear transformations or translations of the wild type gene. In particular, the set of translations that preserve the chemical type of the third base position in the codon is a subgroup which describes the most frequent mutations observed in mutational variants of four genes: human phenylalanine hydroxylase (PAH), human beta globin (HBG), HIV-1 Protease (HIVP) and HIV-1 Reverse transcriptase (HIVRT). Furthermore, an inner pseudo-product defined between codons tends to have a positive value when the codons code to similar amino acids and a negative value when the codons code to amino acids with extreme hydrophobic properties. Consequently, it is found that the inner pseudo-product between the wild type and the mutant codons tends to have a positive value in the mutational variants of the genes: PAH, HBG, HIVP, HIVRT.

[*] Robersy Sánchez: robersy@uclv.edu.cu

Corresponding address: Apartado postal 697. Santa Clara 1. CP 50100. Villa Clara. Cuba

1. **Introduction**

Recently was reported the Boolean lattices of the genetic code (Sánchez et al. 2004a and 2004b). Two dual genetic code Boolean lattices –primal and dual- were obtained as the direct third power of the two dual Boolean lattices $B(X)$ of the four DNA bases: $C(X)=B(X)\times B(X)\times B(X)$. The most elemental properties of the DNA bases and amino acids were used to establish the Boolean lattices $B(X)$ which are isomorphic to $((Z_2)^2, \vee, \wedge)$ and $((Z_2)^2, \wedge, \vee)$ ($Z_2=\{0,1\}$). Consequently, the lattices $C(X)$ are isomorphic to the dual Boolean lattices $((Z_2)^6, \vee, \wedge)$ and $((Z_2)^6, \wedge, \vee)$.

Here, we is used the isomorphism $\varphi: B(X) \to (Z_2)^2$ and the biological importance of base positions in the codons to state a partial order in the codon set and represent the codons as a binary sextuplet. The importance of the base position is suggested by the error frequency found in the codons. Errors on the third base are more frequent than on the first base, and, in turn, these are more frequent than errors on the second base [Woese, 1965; Friedman and Weinstein, 1964; Parker,1989]. These positions, however, are too conservative with respect to changes in polarity of the coded amino acids [Alf-Steinberger, 1969].

The principal aim of this work is to show that a simple Galois field of the genetic code ($C_g$) can be defined on the DNA sequence space allowing us to describe the mutations pathways in the molecular evolution process through the use of the transformations $F: (C_g)^N \to (C_g)^N$, defined on the Galois field of 64 elements ($GF(64)$).

2. **Theoretical Model**

Here, we start from Boolean lattices of the four DNA bases. It is possible develop our theoretical model using both Boolean lattices, primal and dual, but as will see later the primal Boolean lattice leads us to a most significant biological model. So, we will use the binary representation of the four DNA bases of this lattice: G↔00, A↔01, U↔10, C↔11. We have two reasons to use this representation: first, the complementary bases in the DNA molecule are in correspondence with complementary digits and second, this is not an arbitrary base codification, this is the result of an isomorphism between two Boolean lattices, $\varphi: B(X) \to ((Z_2)^2, \wedge, \vee)$, $Z_2 = \{0. 1\}$ (Sánchez et al., 2004a). In addition, to state a correspondence between the codon set and the elements of $GF(64)$, the polynomial representation of the $GF(64)$ will be used(see Appendix).

## 2.1. Nexus between the Galois Field Elements and the Set of Codons

Next, the order of importance of the bases positions in the codons and the isomorphism $\varphi: B(X) \to (Z_2)^2$ allow us to state a function $\Psi: GF(64) \to C_g$, such that:

$$(a_0 + a_1 x + a_2 x^2 + a_3 x^3 + a_4 x^4 + a_5 x^5) \to (f_1(a_2 a_3), f_2(a_4 a_5), f_3(a_0 a_1))$$

The bijective functions $f_k$ have the form:

$$f_k(a_{i(k)} a_{i(k)+1}) = X_k,$$

where $k=1, 2, 3$ denote the codon base position, $i(k) = 2 \cdot k \mod 6$, $a_{i(k)}, a_{i(k)+1} \in \{0,1\}$ and $X_k \in \{G, U, A, C\}$ (or $X_k \in \{C, A, U, G\}$). The functions $f_k$ are equal to the inverse $\varphi^{-1}$ of function $\varphi$ and state the correspondence, in the primal Boolean algebra of the four bases:

$$00 \to G;\ 01 \to A;\ 10 \to U;\ 11 \to C$$

It is not difficult to prove that the function $\Psi$ is bijective, i.e. for all $X_1 X_2 X_3 \in C_d$ there is a polynomial $p(x) \in GF(64)$ and vice verse, such that:

$$\Psi(p(x)) = X_1 X_2 X_3$$

Note that the polynomial coefficients $a_5$ and $a_4$ of the terms with maximal degree, $a_5 x^5$ and $a_4 x^4$ respectively, correspond to the base of second codon position. Next, we found the coefficients that correspond to the first base and finally those of third codon position. That is, the degree of polynomial terms decreases from the most biological important base to the less biologically important base. As a result the ordered codon set showed in Tables 1 are obtained. Note that, in the tables, for every codon its sequence of binary digits is the reverse of the binary digits sequence computed to the corresponding integer number. We have, for instance,

$11 \to 001011 \to 110100\ (AGC) \to 1 + x + x^3$

25→ 011001→ 100110 (AUU) →$1 + x^3 + x^4$

34→ 100010→ 010001 (GAA) →$x + x^5$

## 2.2. Vector Spaces on the Genetic Code Galois Field

Now, by mean of the function $\Psi$ we can define a product operation in the set of codons. Let $\Psi^{-1}$ be the inverse function of $\Psi$ then, for all pair of codons $X_1Y_1Z_1 \in C_g$ and $X_2Y_2Z_2 \in C_g$, their product "·" will be:

$$X_1Y_1Z_1 \cdot X_2Y_2Z_2 = \Psi[\Psi^{-1}(X_1Y_1Z_1)\ \Psi^{-1}(X_2Y_2Z_2)\ mod\ g(x)]$$

That is to say, the product between two codons is obtained from the product of their corresponding polynomials module $g(x)$, where $g(x)$ is an irreducible polynomial of six degree on the $GF(2)$ (see Appendix). Since there are nine irreducible polynomials of six degrees, we have nine possible variant to choose the product between two codons. It is not a problem to prove that the set of codons $(C_g, \cdot)$ with the operation product "·" is an Abelian group. Likewise, we define a sum operation making use the sum operation in $GF(64)$. In this field the sum is carried out by means of the polynomial sum in the usual fashion with polynomial coefficients reduced module 2 (see Appendix).

Then, for all pair of codons $X_1Y_1Z_1 \in C_g$ and $X_2Y_2Z_2 \in C_g$, their sum "+" will be:

$$X_1Y_1Z_1 + X_2Y_2Z_2 = \Psi[\Psi^{-1}(X_1Y_1Z_1) + \Psi^{-1}(X_2Y_2Z_2)]$$

As a result the set of codon $(C_g, +)$ with operation "+" is an Abelian group and the set $(C_g, +, \cdot)$ is a field isomorphic to $GF(64)$. Actually, we have two duals Galois field of codons. After that, we can define the product of a codon $XYZ \in C_g$ by the element $\alpha_i \in GF(64)$. For all $\alpha_i \in GF(64)$ and for all $XYZ \in C_g$, this operation will be defined as:

$$\alpha_i (XYZ) = \Psi[\alpha_i\ \Psi^{-1}(XYZ)\ mod\ 2]$$

**Table 1**. Primal ordered set of codons corresponding to the elements of *GF*(64). In the table is showed the bijection between the codon set and the binary sextuples of $(Z_2)^6$, which are also the coefficients of the polynomials in the *GF*(64) (see Appendix). It is also showed the corresponding integer number of every binary sextuple.

| | G | | | | T | | | | A | | | | C | | |
|---|---|---|---|---|---|---|---|---|---|---|---|---|---|---|---|
| No. | GF(64) | I | II | No. | GF(64) | I | II | No. | GF(64) | I | II | No. | GF(64) | I | II |
| 0 | 000000 | GGG | G | 16 | 000010 | GTG | V | 32 | 000001 | GAG | E | 48 | 000011 | GCG | A |
| 1 | 100000 | GGU | G | 17 | 100010 | GUU | V | 33 | 100001 | GAU | D | 49 | 100011 | GCU | A |
| 2 | 010000 | GGA | G | 18 | 010010 | GUA | V | 34 | 010001 | GAA | E | 50 | 010011 | GCA | A |
| 3 | 110000 | GGC | G | 19 | 110010 | GUC | V | 35 | 110001 | GAC | D | 51 | 110011 | GCC | A |
| 4 | 001000 | UGG | W | 20 | 001010 | UUG | L | 36 | 001001 | UAG | - | 52 | 001011 | UCG | S |
| 5 | 101000 | UGU | C | 21 | 101010 | UUU | F | 37 | 101001 | UAU | Y | 53 | 101011 | UCU | S |
| 6 | 011000 | UGA | - | 22 | 011010 | UUA | L | 38 | 011001 | UAA | - | 54 | 011011 | UCA | S |
| 7 | 111000 | UGC | C | 23 | 111010 | UUC | F | 39 | 111001 | UAC | Y | 55 | 111011 | UCC | S |
| 8 | 000100 | AGG | R | 24 | 000110 | AUG | M | 40 | 000101 | AAG | K | 56 | 000111 | ACG | T |
| 9 | 100100 | AGU | S | 25 | 100110 | AUU | I | 41 | 100101 | AAU | N | 57 | 100111 | ACU | T |
| 10 | 010100 | AGA | R | 26 | 010110 | AUA | I | 42 | 010101 | AAA | K | 58 | 010111 | ACA | T |
| 11 | 110100 | AGC | S | 27 | 110110 | AUC | I | 43 | 110101 | AAC | N | 59 | 110111 | ACC | T |
| 12 | 001100 | CGG | R | 28 | 001110 | CUG | L | 44 | 001101 | CAG | Q | 60 | 001111 | CCG | P |
| 13 | 101100 | CGU | R | 29 | 101110 | CUU | L | 45 | 101101 | CAU | H | 61 | 101111 | CCU | P |
| 14 | 011100 | CGA | R | 30 | 011110 | CUA | L | 46 | 011101 | CAA | Q | 62 | 011111 | CCA | P |
| 15 | 111100 | CGC | R | 31 | 111110 | CUC | L | 47 | 111101 | CAC | H | 63 | 111111 | CCC | P |

This operation is analogous to the multiplication rule of a vector by a scalar. So, $(C_g, +, \cdot)$ can be considered a one-dimensional vector space on *GF*(64). The canonical base of this space is the codon GGU. We shall call this structure the genetic code vector space on *GF*(64). Such structure can be extended to the *N*-dimensional sequence space (*S*) consisting of the set of all $64^N$ DNA sequences with *N* codons. Evidently, this set is isomorphic to the set of all *N*-tuples $(x_1,...,x_N)$ where $x_i \in C_g$. Then, set *S* can be represented by all *N*-tuples $(x_1,...,x_N) \in (C_g)^N$. As a result, the *N*-dimensional vector space of the DNA sequences on *GF*(64) will be the direct sum

$$S = (C_g)^N = C_g \oplus C_g \oplus ... \oplus C_g \text{ (}N\text{ times)}$$

The sum and product in *S* are carried out by components (Redéi, 1967). That is, for all $\alpha \in GF(64)$ and for all $s, s' \in S$ we have:

$$s + s' = (s_1, s_2, ..., s_N) + (s_1', s_2', ..., s_N') = (s_1 + s_1', s_2 + s_2', ..., s_N + s_N')$$

$$\alpha s = \alpha (s_1, s_2, ..., s_N) = (\alpha s_1, \alpha s_2, ..., \alpha s_N)$$

Next, it can proved that $(S, +)$ is an Abelian group with the *N*-tuple $s_e =$ (GGG, GGG,…, GGG) as its neutral element. The canonical base of this space is the set of vectors:

$e_1$=(GGU, GGG, … , GGG), $e_2$=( GGG,GGU,…, GGG), . . . , $e_N$=(GGG, GGG,..., GGU)

As a result, every sequence $s \in S$ has the unique representation as:

$$s = \alpha_1 e_1 + \alpha_2 e_{1+\ldots} + \alpha_N e_N \ (\alpha_i \in GF(64))$$

It is usually said that the *N*-tuple $(\alpha_1, \alpha_2,..., \alpha_N)$ is the coordinate representation of *s* in the canonical bases $\{e_i \in C_g, i=1,2,…,N\}$ of *S*.

In the vector space $C_g$ if we represented the codons as binary sextuplets then the "natural" distance between two codons *X* and *Y* is the Hamming distance $(d_H(X,Y))$. This distance between two codons corresponds to the number of different digits between their binary representations. That is,

$d_H$(CGU, AUC)= $d_H$ (110010, 011011) = 3
$d_H$(AAG, UGA) = $d_H$(010100, 100001) = 4

Next, we shall define in $C_g$ the digital root $r(X_1X_2X_3)$ of a codon $X = X_1X_2X_3$ as the sum of digits in its binary representation. That is, for instance:

$r$(AUG) = $r$(000110) = 2
$r$(CAU) = $r$(101101) = 4

As a result the Hamming distance between two codons *X* and *Y* will be:

$$d_H(X, Y) = r(X + Y)$$

The digital root of one gene will be the binary digits sum of its binary representation.

2.3. Inner pseudo-product in $C_g$ and in $S$

In the $C_g$ we shall define the inner pseudo-product ($\langle X, Y \rangle$) of two codons $X = X_1X_2X_3$ and $Y=Y_1Y_2Y_3$ as:

$$\langle X, Y \rangle = r( X \bullet Y ) - d_H(X, Y ) = r( X \bullet Y ) - r(X + Y ) \quad (1)$$

It is not difficult to see that the inner pseudo-product $\langle X, Y \rangle$ has the following properties:

1) $\langle X, Y \rangle = \langle Y, X \rangle$
2) $\langle X, X \rangle > 0$, for all $X \in C_g$, and $\langle X, X \rangle = 0$ if and only if $X=GGG$.

Property (1) follows due to both the operation product and the Hamming distance are commutative. Property 2 is due to $\langle X, X \rangle = r( X \bullet X ) > 0$, for all $X \neq GGG$, $X \in C_g$. The inner pseudo-product $\langle g_1, g_2 \rangle$ of two DNA sequence $g_1=(c_{11},\ldots, c_{1n})$ and $g_2 = (c_{21},\ldots, c_{2n})$ will be defined as:

$$\langle g_1, g_2 \rangle = r(g_1 \bullet g_2) - r(g_1 + g_2) \quad (2)$$

Since the digital root of a gene is the sum of digital roots of their coordinates we have:

$$\langle g_1, g_2 \rangle = \sum_{i=1}^{n} \langle c_{1i}, c_{2i} \rangle \quad (3)$$

## 3. Results and Discussion

As we see above the Galois field of codons is not unique. Actually, we have obtained nine isomorphic Galois fields, each one with the product operation defined from one of the nine irreducible polynomials. It is convenient, however, to choose a most biologically significant Galois field.

The most attractive irreducible polynomials are the primitive polynomials. If $\alpha_0$ is a root of a primitive polynomial then its powers $\alpha_0^n$ ($n = 1,\ldots, 63$) are the elements of the multiplicative group of $GF(64)$, i.e. $\alpha_0$ is a group generator. Just six of the nine irreducible polynomials are primitives. A common root for all of them is the simplest $\alpha$ (see Table 2). This fact is biologically significant because the element $\alpha_0$ correspond to the codon GGU that code to the simplest amino acid, glycine. From a molecular stand point we can say that glycine structure is present in all amino acids, i.e. glycine has, basically, the structure from which every amino acid is built. In addition, as we show in the Appendix, the product operation in a Galois field generated by a primitive polynomial is carry out in a very simple way.

## 3.1 The Best Biologically Significant Polynomial.

It is expected that some algebraic properties of codons will be connected with the physicochemical properties of amino acids. So, this relationship will allows us to choose one of the six primitive polynomials to define the product operation in $GF(64)$. We expect that, for instance, the difference between algebraic inverse codons will be proportional to the differences between the physicochemical properties of the amino acids coded by them.

In the Boolean lattice of the genetic code it was pointed out a correlation between the mean of Hamming distance ($d_H$) among amino acids –computed from their codons- and the Euclidean distance ($d_E$), stated from their representation as vector of physicochemical properties (Sánchez et al., 2004a).

**Table 2.** Primitive polynomials of six degree on $GF(2)$ and their roots.

| Polynomials | Polynomial roots | | | | | |
|---|---|---|---|---|---|---|
| $1+x+x^6$ | $\alpha$ | $\alpha^2$ | $1+\alpha^3$ | $\alpha^2+\alpha^3$ | $\alpha^4$ | $1+\alpha+\alpha^4$ |
| $1+x+x^3+x^4+x^6$ | $\alpha$ | $\alpha^2$ | $\alpha^4$ | $1+\alpha+\alpha^4$ | $\alpha+\alpha^4+\alpha^5$ | $1+\alpha+\alpha^2+\alpha^4+\alpha^5$ |
| $1+x^5+x^6$ | $\alpha$ | $\alpha^2$ | $\alpha^4$ | $\alpha+\alpha^3+\alpha^4$ | $1+\alpha+\alpha^2+\alpha^5$ | $\alpha+\alpha^3+\alpha^5$ |
| $1+x+x^2+x^5+x^6$ | $\alpha$ | $\alpha^2$ | $\alpha^4$ | $\alpha^2+\alpha^3+\alpha^4$ | $1+\alpha^2+\alpha^4+\alpha^5$ | $\alpha+\alpha^2+\alpha^3+\alpha^4+\alpha^5$ |
| $1+x^2+x^3+x^5+x^6$ | $\alpha$ | $1+\alpha$ | $\alpha^2$ | $1+\alpha^2$ | $\alpha^4$ | $1+\alpha^4$ |
| $1+x+x^4+x^5+x^6$ | $\alpha$ | $\alpha^2$ | $\alpha^4$ | $\alpha+\alpha^3$ | $\alpha+\alpha^3+\alpha^5$ | $1+\alpha+\alpha^2+\alpha^4+\alpha^5$ |

Since the nexus between the Boolean lattice and the Galois field of the genetic code, their metric properties are topologically equivalents. Hence, we can use the distances $d_H$ and $d_E$ to choose the polynomial with the best biological signification. The finest polynomial should produce the best fitting of the equation:

$$d_H = m \, d_E \qquad (4)$$

In this way for every primitive polynomial was computed its multiplicative group in $GF(64)$ and the Hamming distance between the pairs of inverse codons. Next, the Euclidian distance between the pair of amino acids was computed too from their representation as vectors of 12 physicochemical properties. The properties used here are: Mean of area buried on transfer from the standard state to the folded protein, Residue Volume, Normalized van der Waals volume, Polarizability parameter, Polarity, Transfer free energy from octanol to water, Transfer free energy from ciclohexane to water, Transfer free energy from cilooptanol to water, Transfer free energy from vapor to ciclohexane, Transfer free energy surface, Optimized transfer energy parameter, Optimized side chain interaction parameter. These properties were taken from the public database AAindex (http://www.genome.ad.jp/pub/db/genomenet/aaindex/).

Since the numerical scales of all properties are different and ever expressed in different unit the values of all variables were standardized. The measurement employed here was:

$$m_{ij} = (m_{ij} - \mu_{ij})/\sigma_j$$

, where $m_{ij}$ is the raw measurement for amino acid $i$, property $j$; $\mu_{ij}$ the mean of values for the property $j$ over all amino acids and $\sigma_j$ the standard deviation of values for property $j$ over all amino acids.

In the Table 3 the statistical summary of the regression analysis Hamming distance versus Euclidean distance for the six primitive polynomials is shown. The best fitting is obtained with the polynomial $1+x + x^3 + x^4 + x^6$. This polynomial gives us an adjusted R square of 0.87 and fulfills all regression hypotheses. In the Fig 1 the graph of this regression is shown.

**Table 3.** Statistical summary of the regression analysis Hamming distance versus Euclidean distance. The polynomials are represented by means of their coefficients on *GF*(2).

| Primitive Polynomial | Regression Coeficient | Signification | 95% Confidence Interval | | Adjusted R Square | Durbin Watson |
|---|---|---|---|---|---|---|
| | | | Lower Bound | Upper Bound | | |
| 1100111 | 1.152 | 0.000 | 0.940 | 1.364 | 0.79 | 2.090 |
| 1101101 | 1.305 | 0.000 | 1.127 | 1.483 | 0.87 | 1.809 |
| 1110011 | 1.225 | 0.000 | 0.971 | 1.479 | 0.75 | 1.802 |
| 1000011 | 1.225 | 0.000 | 0.995 | 1.455 | 0.79 | 1.548 |
| 1100001 | 1.280 | 0.000 | 1.089 | 1.471 | 0.85 | 2.492 |
| 1011011 | 1.107 | 0.000 | 0.932 | 1.282 | 0.84 | 1.759 |

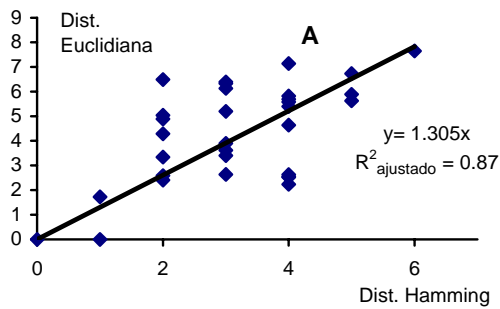

**Figure 1.** Graph of the regression Hamming distance versus Euclidean distance with the polynomial $1+x+x^3+x^4+x^6$.

3.2. Linear Transformations of the DNA Sequences on the *GF*(64).

Gene mutations can be considered linear transformations of the wild type gene in the *n*-dimensional vector space of the DNA sequences. These lineal transformations are the endomorphisms and the automorphisms. In particular, there are some remarkable automorphisms. The automorphism are one-one transformations on the group $(C_g)^N$, such that:

$f(a \cdot (\alpha+\beta)) = a \cdot f(\alpha) + a \cdot f(\beta)$ for all genes $\alpha$ and $\beta$ in $(C_g)^N$ and $a \in GF(64)$

That is, automorphisms forecast mutation reversions, and if the molecular evolution process went by through automorphisms then, the observed current genes do not depend of the mutational pathway followed by the ancestral genes. In addition, the set of all automorphisms is a group.

For every endomorphism (or automorphism) $f: (C_g)^N \to (C_g)^N$, there is a $N \times N$ matrix:

$$A = \begin{pmatrix} a_{11} & \cdots & a_{1N} \\ . & . & . \\ a_{N1} & \cdots & a_{NN} \end{pmatrix}$$

whose rows are the image vectors $f(e_i)$, $i=1,2,...N$. This matrix will be called the representing matrix of the endomorphism $f$, with respect to the canonical base $\{e_i, i=1,2...,N\}$.

In particular, the single point mutations can be considered local endomophisms. An endomorphism $f: S \to S$ will be called local endomorphism if there are $k \in \{1, 2,..., N\}$ and $a_{ik} \in GF(64)$ ($i=1, 2,...,N$) such that:

$$f(e_i) = (0,...,a_{ik},...,0) = a_{ik}e_i$$

This means that:

$$f(x_1, x_2,...x_n) = (x_1, x_2,... \sum_{i=1}^{n} x_i a_{ik},...x_n)$$

It is evident that a local endomorphism will be a local automorphism if, and only if, the elements $a_{kk}$ are different of cero. The endomorphism $f$ will be called diagonal if $f(e_k)=(0,...,a_{kk},...,0)=a_{kk}e_k$ and $f(e_i)=e_i$ for $i \neq k$. This means that:

$$f(x_1,x_2,...x_N) = (x_1,x_2,...a_{kk}x_k,...x_N)$$

The previous concepts allow us to present the following theorem:

**Teorema 1**. For every single point mutation that change the codon $\alpha_i$ of the wild type gene $\alpha = (\alpha_1, \alpha_2,..., \alpha_i,..., \alpha_N)$ ($\alpha$ different of the null vector) by the codon $\beta_i$ of the mutant gene $\beta = (\alpha_1, \alpha_2,...,\beta_i,..., \alpha_N)$, there is:
  i. At least a local endomorphism $f$ such that $f(\alpha) = \beta$.
  ii. At least a local automorphism $f$ such that $f(\alpha) = \beta$.

iii. A unique diagonal automorphism $f$ such that $f(\alpha) = \beta$ if, and only if, the codons $\alpha_i$ and $\beta_i$ of the wild type and mutant genes, respectively, are different of GGG.

Proof: Since for every endomorphism (or automorphism) $f: (C_g)^N \to (C_g)^N$, there is a $N \times N$ matrix $A$ and vice verse then, to prove the theorem it is sufficient to build one endomorphism or automorphism matrix. For every endomorphism $f(\alpha) = (f_1(\alpha),\ldots, f_N(\alpha)) = (\beta_1, \ldots, \beta_N)$ the vector components $f_i(\alpha) = \beta_i$ are the linear combinations of the $i^{th}$-column components of the endomorphism matrix $A$. In particular, for the local endomorphism $f(\alpha) = (f_1(\alpha),\ldots, f_i(\alpha),\ldots, f_N(\alpha)) = (\alpha_1, \ldots, \beta_i,\ldots, \alpha_N)$ it is always possible to build the linear combination:

$$f_i(\alpha) = \sum_{k=1}^{N} \alpha_k a_{ki} = \sum_{\substack{k=1 \\ k \neq i}}^{N} \alpha_k a_{ki} + \alpha_i a_{ii} = \delta_i + \alpha_i a_{ii} = \beta_i$$

In this linear combination the coefficients $a_{ki}$ ($k \neq i$) –of the $i^{th}$-column components of the endomorphism matrix $A$– can be arbitrary chosen and the value of $\delta_i$ fixed. This allows us to solve the equation $\delta_i + \alpha_i a_{ii} = \beta_t$ that always has solution in the $GF(64)$ for $\alpha_i \neq 0$. If $\alpha_i = 0$ we always can fix $\delta_i = \beta_t$. As a result, all $i^{th}$-column components $a_{ki}$ of the endomorphism matrix $A$ are determined, and the remaining components are: $a_{kk} = 1$ and $a_{kl} = 0$ for $k, l \neq i$.

It is sufficient assure $\det(A) \neq 0$ to prove ii. We do it by fixing the coefficient $a_{ii} \neq 0$ to obtain the value $\delta_i$ from the equation $\delta_i + \alpha_i a_{ii} = \beta_t$. Next, the coefficients $a_{ki}$ ($k \neq i$) are arbitrary chosen so that $\sum_{\substack{k=1 \\ k \neq i}}^{N} \alpha_k a_{ki} = \delta_i$

After that, if $\alpha_i \neq$ GGG and $\beta_i \neq$ GGG then chosen the coefficients $a_{ki} = 0$ ($k \neq i$) we have a unique diagonal automorphism because the equation $\beta_i = \alpha_i a_{ii}$ has a unique solution $a_{ii} \neq 0$ and this implies $\det(A) \neq 0$. Conversely, if the local diagonal automorphism is unique then this automorphism leads to the equation $\beta_i = \alpha_i a_{ii}$ that means $\alpha_i \neq$ GGG, $\beta_i \neq$ GGG y $a_{ii} \neq 0$. □

According to the last theorem, any mutation point presented in the Tables 4 and 5, or any combination of these can be represented by means of automorphisms.

Specifically, the most frequent mutation can be described by means of diagonal automorphisms.

### 3.3. Gene Mutations as Translations in *GF*(64).

Gene mutations can be considered translation of the wild type gene in the *N*-dimensional vector space of the DNA sequences. In the Abelian group ($C_g$, +), for two codons $a, b \in (C_g, +)$ the equation $a+x=b$ always has solution then, for all pair of genes $\alpha, \beta \in (C_g, +)^N$ always there is a gene $\kappa \in (C_g, +)^N$ so that $\alpha + \kappa = \beta$. That is, there exists the translation $T: \alpha \rightarrow \alpha + \kappa = \beta$. We shall represent the translation $T_k$ with constant $k$ that act on codon $x$ as:

$$T_k(x) = x + k$$

Any mutation can preserve or change the chemical type of third base in the codon. According to Table 1 for all codon with a pyridine base (C or U) the corresponding integer number of every binary sextuple is an odd number; these codons will be called odd codons. While for all codon with a pyrimidine base (G or A) the corresponding integer number is an even number; these codons will be called even codons. Evidently, those translations with constant k equal an even codon preserve the parity of codons in mutational events. We shall call even translations this kind of translation.

In Tables 4 and 5 we can see that the most frequent mutations keep the codon parity, i.e. they preserve the chemical type of the third base position. Thus the even translation could help us to model the gene mutation process.

Next, we shall consider the composition of translations. Given $W \xrightarrow{f} X \xrightarrow{g} Y$ the composition $g \circ f : W \rightarrow Y$ of translations $g$ and $f$ is defined by $(g \circ f)(x) = g(f(x))$. It is not difficult to see that the set of all translation with composition operation is a group (*G*), and the subset of all even translation $G_T$ is a subgroup of *G*. Next, any mutational pathway followed by genes in the *N*-dimensional vector space will be described by a translation subset of the subgroup $G_T$.

## 3.4. The Inner Pseudo-product and the Physicochemical Properties of Amino Acids

We shall show that the inner pseudo-product is connected with physicochemical properties of amino acids and could help us to understand the gene mutation process. In Table 6 the average of the inner pseudo-product between the codon sets *XAZ*, *XUZ*, *XCZ* and *XGZ* is shown. The most negative values of the inner pseudo-product correspond to the transversions in the second base of codons. It is well-known that such transversions are the most dangerous since they frequently alter the hydrophobic properties and the biological functions of proteins. By contrast, transitions in the second position have the most positive values. In particular, the inner pseudo-product between codons *XAZ* that code to hydrophilic amino acids and codons *XUZ* that code to hydrophobic amino acids have, in general, negative values. This effect is reflected in the average of inner pseudo-products computed for all pairs of amino acids. For the amino acids $a_1$ and $a_2$ with $n$ and $m$ codons the average of inner pseudo-products is computed as:

$$\langle a_1, a_2 \rangle = \frac{1}{n} \sum_{i=1}^{n} \sum_{j=1}^{m} \langle c_{1i}, c_{2j} \rangle \quad (5)$$

The inner pseudo-product average for all amino acid pairs are shown in the Table 7. In general, the inner pseudo-product between amino acids with extreme hydrophobic difference is negative. This is the case, for instance, of the inner pseudo-product average between the hydrophobic amino acids from the set {L, I, M, F, V} and the hydrophilic amino acids from the set {E, D, H, K, N, Q, Y}. Since mutations in genes tend to keep the hydrophobic properties of amino acids, it is natural to think that the inner pseudo-product $\langle c_1, c_2 \rangle$ between codons should be connected with the protein mutation process.

**Table 4.** Mutations in two human genes: beta globin and phenylalanine hydroxylase. The most frequent point mutations are local automorphisms. Beside, the majority of mutations keeps the codon parity and, consequently, can be described as translations (see the text). Those mutations that alter codon parity are in bold type. The inner pseudo-product between the wild type and mutant codons and the absolute difference $|\langle c_W, c_W \rangle_a - \langle c_M, c_M \rangle_a|$ for each gene are written.

| [1]Human Beta Globin | | | | [2]Human PHA | | | |
|---|---|---|---|---|---|---|---|
| [3]Amino Acid Change | Codon Mutation | $\langle c_W, c_M \rangle$ | [4]Diff | Amino Acid Change | Codon Mutation | $\langle c_W, c_M \rangle$ | [2]Diff |
| P36H | CCU-->CAU | 3.00 | 3 | Y204C | UAU-->UGU | 2.00 | 1 |
| T123I | ACC-->AUC | 0.00 | 1 | A104D | GCC-->GAC | 2.00 | 1 |
| V20E | GUG-->GAG | 2.00 | 2 | A165P | GCC-->CCC | 3.00 | 1 |
| V20M | GUG-->AUG | 0.00 | 2 | A246V | GCU-->GUU | 2.00 | 1 |
| V126L | GUG-->CUG | 1.00 | 1 | A259T | GCC-->ACC | 2.00 | 0 |
| V111F | GUC-->UUC | 2.00 | 1 | A259V | GCC-->GUC | 3.00 | 1 |
| **H97Q** | **CAC-->CAA** | 0.00 | 1 | A300S | GCC-->UCC | 1.00 | 1 |
| V34F | GUC-->UUC | 2.00 | 1 | A300V | GCC-->GUC | 3.00 | 1 |
| E121Q | GAA-->CAA | 1.00 | 1 | A309D | GCC-->GAC | 2.00 | 1 |
| L114P | CUG-->CCG | 0.00 | 1 | A309V | GCC-->GUC | 3.00 | 1 |
| A128V | GCU-->GUU | 2.00 | 1 | A313T | CGA-->ACA | -2.00 | 1 |
| **H97Q** | **CAC-->CAG** | 1.00 | 2 | A313V | GCA-->GUA | 3.00 | 3 |
| **D99E** | **GAU-->GAA** | -1.00 | 2 | A322G | GCC-->GGC | 1.00 | 0 |
| D21N | GAU-->AAU | 3.00 | 2 | A322T | GCC-->ACC | 2.00 | 0 |
| N139Y | AAU-->UAU | 2.00 | 3 | A342P | GCA-->CCA | 2.00 | 1 |
| V34D | GUC-->GAC | 3.00 | 0 | A342T | GCA-->ACA | 2.00 | 2 |
| E121K | GAA-->AAA | 1.00 | 0 | A345S | GCU-->UCU | 3.00 | 1 |
| A140V | GCC-->GUC | 3.00 | 1 | A345T | GCU-->ACU | 2.00 | 0 |
| K82E | AAG-->GAG | 1.00 | 0 | A373T | GCC-->ACC | 2.00 | 0 |
| G83D | GGC-->GAC | 4.00 | 1 | A395G | GCC-->GGC | 1.00 | 0 |
| D99N | GAU-->AAU | 3.00 | 2 | A395P | GCC-->CCC | 3.00 | 1 |
| G15R | GGU-->CGU | 1.00 | 1 | A403V | GCU-->GUU | 2.00 | 1 |
| V111L | GUC-->CUC | 1.00 | 1 | A447D | GCC-->GAC | 2.00 | 1 |
| G119D | GGC-->GAC | 4.00 | 1 | A47E | GCA-->GAA | 0.00 | 3 |
| E26K | GAG-->AAG | 1.00 | 4 | A47V | GCA-->GUA | 3.00 | 3 |
| N108I | AAC-->AUC | 0.00 | 0 | C203C | UGC-->UGU | 3.00 | 1 |
| H146P | CAC-->CCC | 3.00 | 2 | C217G | UGU-->GGU | 1.00 | 1 |
| H92Y | CAC-->UAC | 3.00 | 5 | C217R | UGU-->CGU | 3.00 | 0 |
| **C112W** | **UGU-->UGG** | 1.00 | 4 | C265Y | UGC-->UAC | 1.00 | 1 |
| A111V | GCC-->GUC | 3.00 | 1 | C334S | UGC-->UCC | 2.00 | 0 |
| A123S | GCC-->TCC | 1 | 1 | C357G | UGC-->GGC | 1.00 | 1 |

[1]All of the mutation information was taken from the world wide web site: http://globin.cse.psu.edu/ .
[2]All of the mutation information was taken from the *PAHdb* World Wide Web site: http://www.pahdb.mcgill.ca/. [3]The amino acid is represented using the one letter symbol. [4]Diff: Absolute difference: $|\langle c_W, c_W \rangle_a - \langle c_M, c_M \rangle_a|$.

**Table 5.** Mutations in two HIV-1 genes: protease and reverse transcriptase. The most frequent point mutations are local automorphisms. Besides, the majority of mutations keeps the codon parity and, consequently, can be described as translations (see the text). Those mutations that alter codon parity are in bold type. The inner pseudo-product between the wild type and mutant codons and the absolute difference $|\langle c_W, c_W \rangle_a - \langle c_M, c_M \rangle_a|$ for each gene are written.

| [1]Protease | | | | [1]Reverse transcriptase | | | |
|---|---|---|---|---|---|---|---|
| Amino Acid Change | Codon Mutation | $\langle c_W, c_M \rangle$ | [2]Diff | Amino Acid Change | Codon Mutation | $\langle c_W, c_M \rangle$ | [2]Diff |
| A71I | GCU-->AUU | 1.00 | 1 | A62V | GCC-->GUC | 3.00 | 1 |
| A71L | GCU-->CUC | 0.00 | 1 | A98G | GCA-->GGA | 4.00 | 0 |
| A71T | GCU-->ACU | 2.00 | 0 | D67A | GAC-->GCC | 2.00 | 1 |
| A71V | GCU-->GUU | 2.00 | 1 | **D67E** | **GAC-->GAG** | 0.00 | 0 |
| D30N | GAU-->AAU | 3.00 | 2 | D67G | GAC-->GAG | 0.00 | 0 |
| **D60E** | **GAU-->GAA** | -1.00 | 2 | D67G | GAC-->GGC | 4.00 | 1 |
| G16E | GGG-->GAG | -1.00 | 3 | D67N | GAC-->AAC | 3.00 | 2 |
| G48V | GGG-->GUG | -1.00 | 5 | E138A | GAG-->GCG | 0.00 | 1 |
| G52S | GGU-->AGU | 1.00 | 2 | E138K | GAG-->AAG | 1.00 | 0 |
| G73S | GGU-->AGU | 1.00 | 2 | E44A | GAA-->GCA | 0.00 | 3 |
| H69Y | CAU-->UAU | 5.00 | 4 | **E44D** | **GAA-->GAC** | 3.00 | 1 |
| I47V | AUA-->GUA | 2.00 | 2 | E89G | GAA-->GGA | 4.00 | 3 |
| I50L | AUU-->CUU | 1.00 | 1 | E89K | GAA-->GGA | 4.00 | 3 |
| I54L | AUC-->CUC | 3.00 | 1 | F116Y | UUU-->UAU | 2.00 | 2 |
| **I54M** | **AUU-->AUG** | 2.00 | 1 | F77L | UUC-->CUC | 2.00 | 2 |
| I54T | AUC-->ACC | 0.00 | 1 | G141E | GGG-->GAG | -1.00 | 3 |
| I54V | AUC-->GUC | 2.00 | 0 | G190A | GGA-->GCA | 4.00 | 0 |
| I82T | AUC-->ACC | 0.00 | 1 | G190E | GGA-->GAA | 4.00 | 3 |
| I84A | AUA-->GCA | 0.00 | 1 | G190Q | GGA-->CAA | 0.00 | 4 |
| I84V | AUA-->GUA | 2.00 | 2 | G190S | GGA-->UCA | 2.00 | 1 |
| K20M | AAG-->AUG | 3.00 | 0 | G190T | GGA-->ACA | 2.00 | 2 |
| K20R | AAG-->AGG | 2.00 | 1 | G190V | GGA-->GUA | 1.00 | 3 |
| K45I | AAA-->AUA | -1.00 | 2 | G190V | GGA-->GUA | 1.00 | 3 |
| K55R | AAA-->AGA | 3.00 | 1 | G190V | GGA-->GUA | 1.00 | 3 |
| L10I | CUC-->UUC | 2.00 | 2 | H208Y | CAU-->UAU | 5.00 | 4 |
| L10R | CUC-->AUC | 3.00 | 1 | I135M | AUA-->AUG | 2.00 | 1 |
| L10V | CUC-->CGC | 2.00 | 1 | I135T | AUA-->ACA | 2.00 | 1 |
| L10F | CUC-->GUC | 1.00 | 1 | K101Q | AAA-->CAA | 1.00 | 1 |
| L10Y | CUC-->UAC | -1.00 | 2 | K103R | AAA-->AGA | 3.00 | 1 |
| L23I | CUA-->AUA | 3.00 | 1 | K103T | AAA-->ACA | 2.00 | 1 |
| L24I | UUA-->AUA | 1.00 | 1 | K70E | AAA-->GAA | 1.00 | 0 |

[1]All of the mutation information contained in this printed table was taken from the Los Alamos web site: http://resdb.lanl.gov/Resist_DB.. [2]Diff: Absolute difference: $|\langle c_W, c_W \rangle_a - \langle c_M, c_M \rangle_a|$.

In accordance with Tables 6 and 7, in the most frequent codon mutations observed in genes, the inner pseudo-product between the wild type and the mutant codons should be a positive value.

The inner pseudo-product between the wild type and mutant codons in mutational variants of two human genes: beta globin and phenylalanine hydrolase are shown in Table 4. In most frequent mutations the inner pseudo-product values are greater than -1. A similar situation is found in two HIV-1 genes: protease and reverse transcriptase (Table 5).

In addition, it is found that the magnitude $\langle a, a \rangle$ tends to rise with the increase of the average of volume buried of amino acids $V_b$ in proteins (Chothia, 1975). A similar tendency it is found between $\langle a, a \rangle$ and the mean of area buried on transfer from standard state to the folded protein ($A_b$) (Rose, et al., 1985). The best fit –excluding amino acid Triptophan– leads us to the equations:

$$\langle a, a \rangle = 0.0148\ V_b\ (R^2_{adjusted} = 0.86) \quad (6)$$
$$\langle a, a \rangle = 0.0167 A_b\ (R^2_{adjusted} = 0.84) \quad (7)$$

So the inner pseudo-product is associated with topological variables that express the degree to which amino acid residues are buried by backbone atoms from covalent neighbors in the folded protein. In another way, as might be expected the variables $V_b$ and $A_b$ are proportional to the molecular weight of amino acids ($MW$). So we have the expression:

$$V_b = 0.891\ MW \quad (8)\ (R^2_{adjusted} = 0.99)$$
$$A_b = 0.969\ MW \quad (9)\ (R^2_{adjusted} = 0.969)$$

**Table** 6. The average of inner pseudo-product between codon subsets $XAZ$, $XUZ$, $XCZ$ and $XGZ$. Behind each codon subset, for example, $XUZ$ there are 16 realizations. Thus, for every pair of codon subsets there is a symmetric distance matrix with 162 elements. The inner pseudo-product between two codon subsets is the mean of the 256 inner pseudo-products between their codons.

|     | XGZ    | XUZ    | XAZ    | XCZ    |
|-----|--------|--------|--------|--------|
| XGZ | 0.531  | -0.031 | -0.094 | -1.156 |
| XUZ | -0.031 | 0.969  | -0.969 | 0.031  |
| XAZ | -0.094 | -0.969 | 1.031  | 0.031  |
| XCZ | -1.156 | 0.031  | 0.031  | 1.094  |

**Table 7**. Average of inner pseudo-product for all amino acid pairs. The negative values of inner pseudo-product are in bold type. For amino acid glycine the codon GGG was not considered.

|   | G | W | C | R | S | V | L | F | M | I | E | D | Y | K | N | Q | H | A | T | P |
|---|---|---|---|---|---|---|---|---|---|---|---|---|---|---|---|---|---|---|---|---|
| G | 0.67 | **-1.00** | 0.83 | **-0.17** | **-0.17** | 0.50 | **-0.39** | 1.17 | **-1.33** | 0.67 | 1.50 | 1.50 | 0.17 | **-0.17** | **-0.17** | **-0.83** | **-0.83** | 0.50 | **-0.50** | **-1.17** |
| W | **-1.00** | 1.00 | 1.00 | 0.33 | 0.33 | 1.00 | 1.00 | 1.00 | 2.00 | 0.67 | 2.00 | 0.00 | 0.00 | 0.00 | **-2.00** | 0.00 | **-2.00** | **-1.00** | **-3.00** | **-1.00** |
| C | 0.83 | 1.00 | 2.75 | 0.75 | 0.58 | **-0.50** | 0.58 | 1.25 | **-1.50** | **-0.83** | **-0.75** | **-0.25** | 1.25 | **-0.25** | 0.25 | **-0.75** | 0.75 | **-1.00** | **-1.50** | **-1.50** |
| R | **-0.17** | 0.33 | 0.75 | 2.08 | **-0.58** | **-0.58** | 0.53 | **-0.58** | 1.17 | 0.17 | **-0.42** | **-0.42** | **-0.42** | 0.58 | 0.58 | 0.92 | 0.25 | **-1.92** | **-0.58** | **-0.25** |
| S | **-0.17** | 0.33 | 0.58 | **-0.58** | 0.14 | 0.33 | **-0.19** | 0.58 | **-1.50** | 0.06 | **-0.42** | 0.42 | 1.25 | **-0.75** | 0.08 | **-0.25** | 0.58 | 0.33 | 0.00 | 0.50 |
| V | 0.50 | 1.00 | **-0.50** | **-0.58** | 0.33 | 2.13 | 0.00 | 1.00 | 0.50 | 1.00 | 0.25 | 0.00 | **-0.25** | **-1.75** | **-0.50** | **-1.50** | **-2.75** | 0.88 | 0.13 | **-0.88** |
| L | **-0.39** | 1.00 | 0.58 | 0.53 | **-0.19** | 0.00 | 1.64 | 1.42 | 1.83 | 0.50 | **-1.42** | **-1.92** | **-0.75** | **-1.08** | **-1.92** | 0.08 | **-0.42** | **-0.50** | **-0.33** | 0.67 |
| F | 1.17 | 1.00 | 1.25 | **-0.58** | 0.58 | 1.00 | 1.42 | 1.75 | **-1.50** | 0.50 | **-1.75** | **-0.25** | 0.25 | **-1.75** | **-2.25** | **-1.25** | **-0.75** | 0.00 | **-1.00** | 0.00 |
| M | **-1.33** | 2.00 | **-1.50** | 1.17 | **-1.50** | 0.50 | 1.83 | **-1.50** | 3.00 | 1.00 | **-1.50** | **-1.50** | **-3.50** | 2.50 | 0.50 | **-1.50** | **-1.50** | 0.50 | 1.50 | 0.50 |
| I | 0.67 | 0.67 | **-0.83** | 0.17 | 0.06 | 1.00 | 0.50 | 0.50 | 1.00 | 2.33 | **-1.17** | **-0.83** | **-2.50** | **-0.50** | 0.50 | **-0.83** | **-1.17** | **-0.33** | 1.33 | **-0.33** |
| E | 1.50 | 2.00 | **-0.75** | **-0.42** | **-0.42** | 0.25 | **-1.42** | **-1.75** | **-1.50** | **-1.17** | 2.75 | 0.75 | 1.25 | 1.25 | 1.25 | 0.75 | **-1.25** | 1.25 | **-0.25** | **-0.75** |
| D | 1.50 | 0.00 | **-0.25** | **-0.42** | 0.42 | 0.00 | **-1.92** | **-0.25** | **-1.50** | **-0.83** | 0.75 | 2.25 | 1.25 | 1.25 | 1.75 | **-0.75** | 0.75 | 1.00 | 0.00 | **-1.50** |
| Y | 0.17 | 0.00 | 1.25 | **-0.42** | 1.25 | **-0.25** | **-0.75** | 0.25 | **-3.50** | **-2.50** | 1.25 | 1.25 | 1.75 | **-0.25** | **-0.25** | 0.25 | 3.25 | **-0.25** | **-0.75** | **-0.75** |
| K | **-0.17** | 0.00 | **-0.25** | 0.58 | **-0.75** | **-1.75** | **-1.08** | **-1.75** | 2.50 | **-0.50** | 1.25 | 1.25 | **-0.25** | 2.25 | 0.25 | 2.75 | 0.75 | 0.25 | 0.25 | 0.75 |
| N | **-0.17** | **-2.00** | 0.25 | 0.58 | 0.08 | **-0.50** | **-1.92** | **-2.25** | 0.50 | 0.50 | 1.25 | 1.75 | **-0.25** | 0.25 | 2.75 | 0.25 | 1.75 | 0.00 | 1.00 | **-0.50** |
| Q | **-0.83** | 0.00 | **-0.75** | 0.92 | **-0.25** | **-1.50** | 0.08 | **-1.25** | **-1.50** | **-0.83** | 0.75 | **-0.75** | 0.25 | 2.75 | 0.25 | 3.75 | 1.25 | **-1.00** | **-0.50** | 0.50 |
| H | **-0.83** | **-2.00** | 0.75 | 0.25 | 0.58 | **-2.75** | **-0.42** | **-0.75** | **-1.50** | **-1.17** | **-1.25** | 0.75 | 3.25 | 0.75 | 1.75 | 1.25 | 4.25 | **-1.25** | 0.25 | 1.75 |
| A | 0.50 | **-1.00** | **-1.00** | **-1.92** | 0.33 | 0.88 | **-0.50** | 0.00 | 0.50 | **-0.33** | 1.25 | 1.00 | **-0.25** | 0.25 | 0.00 | **-1.00** | **-1.25** | 2.38 | 1.38 | 0.13 |
| T | **-0.50** | **-3.00** | **-1.50** | **-0.58** | 0.00 | 0.13 | **-0.33** | **-1.00** | 1.50 | 1.33 | **-0.25** | 0.00 | **-0.75** | 0.25 | 1.00 | **-0.50** | 0.25 | 1.38 | 2.13 | 0.88 |
| P | **-1.17** | **-1.00** | **-1.50** | **-0.25** | 0.50 | **-0.88** | 0.67 | 0.00 | 0.50 | **-0.33** | **-0.75** | **-1.50** | **-0.75** | 0.75 | **-0.50** | 0.50 | 1.75 | 0.13 | 0.88 | 1.88 |

Now, from the equality (3) and equations (8) or (9) it follows:

$$\langle g, g \rangle = \sum_{i=1}^{n} \langle c_i, c_i \rangle \approx \sum_{i=1}^{n} \langle a_i, a_i \rangle = \langle g, g \rangle_a = \beta \sum_{i=1}^{n} MW_i \approx \beta\, MW_p \qquad (10)$$

where, the inner pseudo-product of every codon $\langle c_i, c_i \rangle$ is replaced by the average of inner pseudo-product for all corresponding synonymous codons $\langle a_i, a_i \rangle$ and the sum $\sum_{i=1}^{n} MW_i$ of the amino acid molecular weight is replaced by protein molecular weight (to form every peptide linkage of a polypeptide chain a water molecular is lost). The lineal regression analysis with 471 proteins, taken from the protein data bank, confirms the last equality given to us by the expression:

$$\langle g, g \rangle_a = 18.30\, MW_p \qquad (11)\ (R^2_{adjusted} = 0.999)$$

The graph of this regression is shown in Fig. 2.

It has been pointed out by Chotia that protein interiors are closely packed, each residue occupying the same volume as it does in crystals of amino acids (Chothia, 1975).

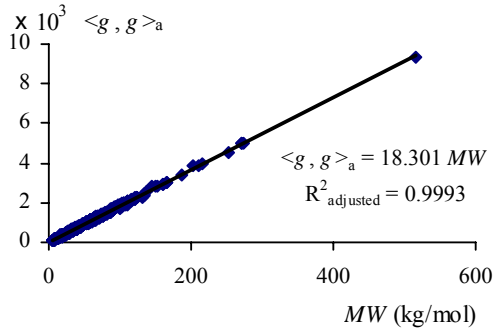

**Figure 2**. Graph of the regression analysis of the inner pseudo-product $\langle g, g \rangle_a$ versus the molecular weight $MW$. The 95% Confidence Interval for regression coefficient is: Lower bound, 18.258, and upper bound 18.344.

As a result, allowing for equation (6) and (7), in the gene mutation process we should expect a small change of inner pseudo-product of codons, i.e. we should expect a small value of the absolute difference $|\langle c_W, c_W \rangle_a - \langle c_M, c_M \rangle_a|$ between the inner pseudo-product of the wild type and mutant codons. Such result is confirmed in Tables 4 and 5 where the most frequent values are close to 1.

The inner pseudo-product reflects the quantitative relationships between codons in genes. These relationships are suggested by the codons usage found in genes (Nakamura, et al., 2001). In all living organisms, note that some amino acids and some codons are more frequent than others (see http://www.kazusa.or.jp/codon). Each organism has its own "preferred" or more frequently used codons for a given amino acid and their usage is frequent, a tendency called codon bias. For all life forms, codon usage is non-random (Fuglsang, 2003) and associated to various factors such as gene expression level (Makrides, 1996), gene length (Duret and Mouchiroud, 1999) and secondary protein structures (Oresic and Shalloway, 1998; Tao and Dafu, 1998; Gupta et el., 2000). Moreover, most amino acids in all species bear a highly significant association with gene functions, indicating that, in general, codon usage at the level of individual amino acids is closely coordinated with the gene function (Fuglsang, 2003). So, the constraints observed in the values of $\langle c_W, c_M \rangle$ and $|\langle c_W, c_W \rangle_a - \langle c_M, c_M \rangle_a|$ in Tables 4 and 5 are consequence of the codons usage which restrict the number of mutational variants in point mutations in genes. As a result, the connection between codon usage and protein structure leads to the relationship between the inner pseudo product and the topological variables $V_b$ and $A_b$.

## 4. Conclusions

The isomorphism between the Boolean lattice of the four DNA bases and the Boolean lattice $((Z_2)^2, \wedge, \vee)$ allows us to define a new Galois field of the genetic code. On this new field it was defined a new *N*-dimensional DNA sequence vector space where gene mutations can be considered linear transformation or translation of the wild type gene. It is proved that for every single point mutation in the wild type gene there is at least an automorphism that transforms the wild type in the mutant gene. Besides this, it is found that the set of translations that preserves the chemical type of the third base position in the codon is a subgroup which describes the most frequent mutation observed in mutational variants of four genes: PAH, HBG, VP and VRT.

The inner pseudo-product $\langle c_1, c_2 \rangle$ defined between codons showed strong connection with the hydrophobic properties of amino acids. This product tends to have a positive value between similar amino acids and a negative value between amino acids with extreme hydrophobic properties. As a result, we should expect that the most frequent values of the inner pseudo-product between the wild type and the mutant codons in gene mutation process should be positive values. This fact is confirmed in the four mutational variants of genes: PAH, HBG, VP and VRT.

In addition, the average of the inner pseudo-product $\langle a, a \rangle_a$ for every amino acid has a lineal correlation with the volume and area of amino acids buried in the folded protein. Due to the tendency of gene mutation to keep the protein structure it is expected that the difference between inner pseudo-products of wild type and mutant codons $|\langle a_w, a_w \rangle_a - \langle a_M, a_M \rangle_a|$ should be small. Like to the previous results this tendency was confirmed in the above mentioned four genes.

Finally, it was found that there is an strong lineal correlation between the inner pseudo-product $\langle g, g \rangle_a$ of genes and the their molecular weight.

## Appendix

For the usefulness of the reader, in this appendix we review the definitions of group, field and Vector space. The basic ideas were taken from the books [4, 15, 18]. Besides, we have written a summary about the operation sum and product in $GF(64)$.

**Definition:** A binary operation on $S$ is a function from $S \times S$ to $S$.

In other words a binary operation on $S$ is given when to every pair $(x, y)$ of elements of $S$ another element $z \in S$ is associated. If "$\bullet$" is the binary operation on $S$, then $\bullet(x, y)$ will be denoted by $x \bullet y$, that is the image element $z$ is denoted by $x \bullet y$.

**Definition:** A group is the pair $(G, \bullet)$ composed by a set of elements G and the binary operation $\bullet$ on $G$, which for all $x, y, z \in G$ satisfies the following laws:

i. *Associative law*: $(x \bullet y) \bullet z = x \bullet (y \bullet z)$
ii. *Identity law*: There exists in G a neutral element $e$ such that: $x \bullet e = e \bullet x$
iii. *Inverse law*: For all element $x$ there is the symmetric element $x^{-1}$ respect to $e$ such that:

$x \bullet x^{-1} = x^{-1} \bullet x = e$ (the element $e$ is called neutral element of $G$)

In particular, the subset $H \subset G$ is called a subgroup in $G$ if $e \in G$; $h_1, h_2 \in H \Rightarrow h_1 \bullet h_2 \in H$ and $h \in H \Rightarrow h^{-1} \in H$. Besides, the group $(G, \bullet)$ is called an Abelian group (a commutative group) if for all $x, y \in G$ the binary operation satisfies the commutative law: $x \bullet y = y \bullet x$. For the Abelian group the binary operation is denoted by the symbol "+" and it is called sum operation. Now, the symbol 0 denotes the neutral element.

**Definition.** A field is a set $F$ with two binary operations, denoted by "+" and "$\bullet$", with the following properties:

i. $(F, +)$ is a commutative group
ii. $(F, \bullet)$ is a commutative group
iii. The following holds:

$$(x + y) \bullet z = x \bullet z + y \bullet z$$
$$z \bullet (x + y) = z \bullet x + z \bullet y.$$

**Definition.** Let *F* be field and let *V* be an Abelian group. *V* is called a vector space on the field *F* is there exists an external law $f: F \times V \to V$, given for $f(x,u) = x\, u = u\, x$ that has, for all $x, y \in F$ and for all $u, v \in V$ the following properties:

1. $x(u + v) = xu + xv$
2. $(x + y)v = xv + yv$
3. $(x \bullet y)v = x(yv)$
4. $1\, v = v$

*Galois Field Operations Summary*

Here we used the polynomial representation of the Galois field *GF*(64). This representation is obtained from the quotient ring $F[x]/(g(x))$:

$$h(x) \bmod g(x)$$

where, $F[x]$ is a polynomial set on the field *GF*(2), $h(x) \in F[x]$ and $g(x)$ is an irreducible polynomial of six degree on GF(2). From the finite field theory it is known that the ring $F[x]/(g(x))$ is a finite field representative of the *GF*(64).

In *GF*(64) the sum operation is carried out by means of the polynomial sum in the usual fashion with polynomial coefficients reduced module 2, while the product is the polynomial product module $g(x)$. That is, for all $p_1(x), p_2(x) \in F[x]/(g(x))$, we have:

$$p_1(x) + p_2(x) \bmod 2 = p(x) \in F[x]/(g(x))$$
$$p_1(x) \cdot p_2(x) \bmod g(x) = q(x) \in F[x]/(g(x))$$

For instance, on *GF*(2) the polynomial $1+t^5+t^6$ is an irreducible polynomial and we have:

$(1+t^3) + (t+t^3) \bmod 2 = t + 1$
$(1+ t + t^2)(1+ t) \bmod (1+t^5+t^6) = t^3 + 1$
$(t+t^2 +t^4 + t^5)(1+ t^2+t^3+t^5) \bmod (1+t^5+t^6) = t + t^3$

The expression $p(x) \bmod g(x))$ is the polynomial remainder obtained from the division of $p(x)$ by $g(x)$ according to the Euclidean algorithm for polynomial division.

It can be noted that for every integer number there is a binary representation that leads to polynomial coefficients. We have for instance:

| Integer number | Binary representation | Polynomial coefficients | Polynomial |
|---|---|---|---|
| s = 11 | 1011 | 110100 | $1 + x + x^3$ |
| s = 13 | 1101 | 101100 | $1 + x^2 + x^3$ |
| s = 25 | 11001 | 100110 | $1 + x^3 + x^4$ |
| s = 34 | 100010 | 010001 | $x + x^5$ |

That is to say, there is a bijective function $f[s]$ such that $f: s \to GF(64)$, between the subset of the integer number $s = \{0, 1,\ldots, 63\}$ and the elements of $GF(64)$. According to the above example $f[11] = 1 + x + x^3$, $f[13] = 1 + x^2 + x^3$, $f[25] = 1 + x^3 + x^4$ and $f[34] = x + x^5$.

In the $GF(64)$ one element is called primitive if for all $x \in GF(2^6)$, $x \neq 0$ we have $x = \alpha^i$, where $i \in \{0, 1,\ldots, 63\}$. If the irreducible polynomial $g(x)$ has a root which is a primitive element of $GF(64)$ then $g(x)$ is called primitive polynomial. In a Galois field generated by primitive polynomial it is very ease to carry out the product between two elements. In this field any root of the primitive polynomial is a generator of the multiplicative group. This fact suggests the definition of a logarithm function. If $\alpha$ is a primitive root of the polynomial $g(x)$, we shall call it logarithm base $\alpha$ of the element $\beta$ to the number $n$ for which holds the equality:

$$\alpha^n \bmod g(x) = \beta$$

Now, we can write:

$$f[s] = \alpha^n \bmod p(\alpha)$$

And

$$n = \text{logaritmo}_\alpha f[s] = \log_\alpha f[s]$$

The properties of this logarithm function are alike to the classical definition in Arithmetic:

i. $\log_\alpha (f[x]*f[y]) = (\log_\alpha f[x] + \log_\alpha f[y]) \bmod 63 = (n_x + n_y) \bmod 63$
ii. $\log_\alpha (f[x]/f[y]) = (\log_\alpha f[x] - \log_\alpha f[y]) \bmod 63 = (n_x - n_y) \bmod 63$
iii. $\log_\alpha f[x]^m = m \log_\alpha f[x] \bmod 63$

The logarithm table for the primitive polynomial $1+ x + x^3 + x^4 + x^6$ is shown in the Table 1. We can compute, for instance:

$f[34]*f[21] \rightarrow \log_\alpha (f[34]*f[21]) = \log_\alpha f[34] + \log_\alpha f[21] \bmod 63 = (36 + 40) \bmod 63 = 13$

after that, according to Table 1:

$$f[34] * f[21] = f[9]$$

**Table 1**. Logarithm table of the elements of the GF(64) generated by the primitive polynomial $g(x) = 1+ x + x^3 + x^4 + x^6$. Here, the primitive root $\alpha$ is the simplest root $x$, i.e. $f[s] = x^n \bmod g(x)$ and $n$ = logarithm base $\alpha$ of $f[s] = \log_\alpha f[s]$.

| Element | f[1] | f[2] | f[3] | f[4] | f[5] | f[6] | f[7] | f[8] | f[9] | f[10] | f[11] |
|---|---|---|---|---|---|---|---|---|---|---|---|
| *n* | 0 | 1 | 56 | 2 | 49 | 57 | 20 | 3 | 13 | 50 | 53 |
| **Element** | f[12] | f[13] | f[14] | f[15] | f[16] | f[17] | f[18] | f[19] | f[20] | f[21] | f[22] |
| *n* | 58 | 25 | 21 | 42 | 4 | 35 | 14 | 16 | 51 | 40 | 54 |
| **Element** | f[23] | f[24] | f[25] | f[26] | f[27] | f[28] | f[29] | f[30] | f[31] | f[32] | f[33] |
| *n* | 18 | 59 | 31 | 26 | 6 | 22 | 46 | 43 | 37 | 5 | 30 |
| **Element** | f[34] | f[35] | f[36] | f[37] | f[38] | f[39] | f[40] | f[41] | f[42] | f[43] | f[44] |
| *n* | 36 | 45 | 15 | 34 | 17 | 39 | 52 | 12 | 41 | 24 | 55 |
| **Element** | f[45] | f[46] | f[47] | f[48] | f[49] | f[50] | f[51] | f[52] | f[53] | f[54] | f[55] |
| *n* | 62 | 19 | 48 | 60 | 10 | 32 | 28 | 27 | 9 | 7 | 8 |
| **Element** | f[56] | f[57] | f[58] | f[59] | f[60] | f[61] | f[62] | f[63] | | | |
| *n* | 23 | 11 | 47 | 61 | 44 | 29 | 38 | 33 | | | |